\documentclass[11pt]{article}
\usepackage{amsfonts,latexsym,epic}

\makeatletter
\@addtoreset{equation}{section}
\makeatother
\renewcommand{\theequation}{\thesection.\arabic{equation}}

\newfont{\blackb}{msbm10 scaled\magstep1}
 
\newcounter{subequation}[equation]
\makeatletter

\expandafter\let\expandafter
\reset@font\csname reset@font\endcsname

\def\subeqnarray{\arraycolsep1pt
    \def\@eqnnum\stepcounter##1{\stepcounter{subequation}%
        {\reset@font\rm(\theequation\alph{subequation})}}
\jot5mm     \eqnarray}

\def\subarray{\arraycolsep1pt
    \def\@eqnnum\stepcounter##1{\stepcounter{subequation}%
        {\reset@font\rm(\alph{subequation})}}
\jot5mm     \eqnarray}

\makeatother

\newfont{\calig}{cmsy10 scaled\magstep1}

\def\text#1{\hbox{#1}}
 
\newtheorem{theorem}{Theorem}[section]

\newtheorem{remark}{Remark}[section]

\newcommand{\qed}{\hfill$\Box$}

\def\nn{\nonumber}
\def\non{\nonumber\\}
\def\be{\begin{equation}}
\def\ee{\end{equation}}
\def\ben{\begin{displaymath}}
\def\een{\end{displaymath}}
\def\baa{\begin{eqnarray}}                           
\def\ea{\end{array}}
\def\eaa{\end{eqnarray}} 
\def\ba{\begin{array}}
\def\ea{\end{array}}

\def\g{\gamma}

\def\3{\ss}

\def\d{\delta}
\def\e{\varepsilon}

\def\ka{\kappa}
\def\l{\lambda}

\def\s{\sigma}
\def\t{\tau}
\def\th{\vartheta}

\def\phi{\varphi}

\def\cO{{\cal{O}}}

\def\Zf{{\cal Z}}

\def\C{\mathbb{C}}
\def\Z{\mathbb{Z}}

\def\t0{\Theta_0}

\def\la{\label}
\def\Ref{\ref}
\def\c{\cite}
\def\f{\frac}

\def\p{\partial}

\def\Ref#1{(\ref{#1})}
\def\tr{{\rm tr}}
\def\0{S}
\def\1{T}

\def\log{\ln}

\def\res{{\rm res}}
\def\hPsi{{\widehat{\Psi}}}
\def\htau{{\widehat{\tau}}}
\def\tht{{\widehat{t}}}
\def\hA{{\widehat{A}}}
\def\hG{{\widehat{G}}}
\def\hH{{\widehat{H}}}
\def\hM{{\widehat{M}}}
\def\i{{\rm i}}
\def\sl{{\mathfrak{sl}}}
\def\gl{{\mathfrak{gl}}}

\def\ft#1#2{{\textstyle {\frac{#1}{#2}} }}

\def\A{{\mbox{\sc \footnotesize a}}}
\def\Ab{{\mbox{\sc \footnotesize b}}}
\def\Ac{{\mbox{\sc \footnotesize c}}}

\begin{document}
\begin{flushright}
AEI-1999-31\\
LPTENS-99/36\\
solv-int/9910010
\end{flushright}
\vskip0.3cm
\begin{center}
{\bf \LARGE Schlesinger transformations\\[6pt] 
for elliptic isomonodromic deformations}
\vskip0.6cm
 
{\bf\large D.~Korotkin\,$^a$, N.~Manojlovi\'c\,$^b$, H.~Samtleben\,$^c$}\\
\vskip0.5cm
$^a$Max-Planck-Institut f\"ur Gravitationsphysik,\\
 Am M\"uhlenberg  1, D-14476 Golm, Germany\\
{\small E-mail: korotkin@aei-potsdam.mpg.de}\\[4pt]
$^b$\'Area Departmental de Matem\'atica, U.C.H.E., Campus de Gambelas,\\ 
Universidade do Algarve, 8000 Faro, Portugal\\
{\small E-mail: nmanoj@mozart.si.ualg.pt }\\[4pt]
$^c$ Laboratoire de Physique Th{\'e}orique de l' Ecole Normale
Sup{\'e}rieure,\\  
24 Rue Lhomond, 75231 Paris Cedex 05, France
\footnote{Unit{\'e} Mixte de Recherche du Centre
National de la Recherche Scientifique et de l'Ecole Normale
Sup{\'e}rieure.} \\   
{\small E-mail: henning@lpt.ens.fr}
\end{center} 
\vskip0.6cm
 
\begin{abstract}
Schlesinger transformations are discrete monodromy preserving symmetry
transformations of the classical Schlesinger system. Generalizing
well-known results from the Riemann sphere we construct these
transformations for isomonodromic deformations on genus one Riemann
surfaces. Their action on the system's tau-function is computed and
we obtain an explicit expression for the ratio of the old and the
transformed tau-function. 
\end{abstract} 

\begin{center}
\end{center}
\vskip0.6cm

\section{Introduction}

The theory of isomonodromic deformations of ordinary matrix
differential equations of the type 
\be\la{A} \f{d\Psi}{d\l}= A(\l)\,\Psi
\;, 
\ee 
where $A(\l)$ is a matrix-valued meromorphic function on
$\overline{\C}$, is a classical area intimately related to the matrix
Riemann-Hilbert problem on the Riemann sphere.  Over the last 20 years
this has become a powerful tool in areas like soliton theory,
statistical mechanics, theory of random matrices, quantum field theory
etc. The main object associated with the isomonodromic deformation
equations is the so-called $\tau$-function which turns out to be
closely related to the Fredholm determinant of certain integral
operators associated to the Riemann-Hilbert problem. After the
classical work of Schlesinger \cite{Schl12} the major contributions to
the development of the subject were made in the papers of Jimbo, Miwa
and their collaborators in the early 80's
\cite{JiMiMoSa80,JiMiUe81,JimMiw81b,JimMiw81c}.

There are only a few cases where the matrix Riemann-Hilbert problem
may be solved explicitly in terms of known special functions (in
addition to the mentioned papers of the Kyoto school see also the
recent work \cite{KitKor98,DIKZ99} where certain classes of solutions
were obtained using the theory of theta-functions). However, as was
already discovered by Schlesinger himself, there exists a large class
of transformations which allow to get an infinite chain of new
solutions starting from the known ones. They share the characteristic
feature that they shift the eigenvalues of the residues of the
connection $A(\l)$ in \Ref{A} by integer or half-integer values, thus
changing the associated monodromies by sign only. These
transformations -- nowadays called Schlesinger transformations -- were
systematically studied in \cite{JimMiw81b,JimMiw81c}. In particular,
it turns out that being written in terms of the $\tau$-functions the
superposition laws of these transformations provide a big supply of
discrete integrable systems.

The natural question of generalizing the theory of isomonodromic
deformations on the sphere to higher genus surfaces was addressed by
several authors. Here, we mention the contributions of Okamoto
\cite{Okam71,Okam79} and Iwasaki \cite{Iwas91}.

For the case of the torus, recently two different explicit forms of
isomonodromic deformations were proposed. In \cite{KorSam97}, two of
the present authors studied isomonodromic deformations of
non-singlevalued meromorphic connection on the torus whose ``twists''
(which determine the transformation of the connection $A(\l)$ with
respect to tracing along basic cycles of the torus) vary with respect
to the deformation parameters. The isomonodromic deformation equations
for these connections hence contain transcendental dependence on the
dynamical variables, which makes it difficult to analyse this system
in a way analogous to the Schlesinger system on the sphere.  On the
other hand, Takasaki \cite{Taka98} considered connections on the torus
whose twists remain invariant with respect to the parameters of
deformation. In Takasaki's form, the equations of isomonodromic
deformations have already the same degree of non-linearity as the
ordinary Schlesinger system.

The purpose of the present paper is the extension of the notion of
Schlesinger transformations from the Riemann sphere to the
isomonodromy deformation equations on the torus with constant twists
(we call this the elliptic Schlesinger system). In particular, in
analogy to the ordinary Schlesinger system, it turns out to be
possible to derive the action of elliptic Schlesinger transformation
on the $\tau$-function. Thereby, we realize the first steps of the
program to extend the results of \cite{JiMiUe81,JimMiw81b,JimMiw81c}
to the elliptic case. Throughout, we restrict to the case $A(\l)\in
\sl(2,\C)$.

The paper is organized as follows. In section 2 we remind the
framework of isomonodromic transformations on the sphere and reproduce
some facts about the Schlesinger transformations on the Riemann sphere
in a form convenient for generalization to the elliptic case.  In
section 3 we describe the elliptic Schlesinger system with constant
twists. In particular, we find the simple formula
\be
H_{\mu}=- \f{1}{2\pi \i}\oint_a\; \tr A^2(\l)d\l \;,
\ee
for the Hamiltonian which generates the isomonodromic evolution with
respect to the module $\mu$ of the torus. This evolution is hence of
the same type as the isomonodromic evolution with respect to the
position of the singularities $\l_j$ of $A(\l)$, which is generated by
the contour integrals
\be
H_{j}= \f{1}{4\pi \i}\oint_{\l_j} \tr A^2(\l)d\l \;.
\ee
Generalizing the construction from the Riemann sphere we subsequently
implement the elliptic Schlesinger transformations. 

Finally, in section 4 we determine the action of the elliptic
Schlesinger transformations on the $\tau$-function of the system. The
transformed $\tau$-function $\htau$ differs from the old one by a
factor which may be explicitly integrated in terms of the
characteristic parameters of the solution $\Psi$ of \Ref{A}. For
elementary elliptic Schlesinger transformations which shift the
eigenvalues of the residues of $A(\l)$ in two singularities $\l_k$ and
$\l_l$ by $1/2$, the main result is given by Theorem 4.1 below:
\be
\htau\left(\{\l_j\},\mu\right)~=~ 
\left\{\Big[w_1 w_2 w_3\!\left(\f{\l_k\!-\!\l_l}{2}\right)\; 
\Big]^{1/2}\;\det \Big[G J^{1/2}\Big]\:\right\}\,
\cdot\, \tau\left(\{\l_j\},\mu\right) \;,
\ee
with certain elliptic functions $w_\A$, and where $J$ and $G$
are parameters in the local expansion of the solution $\Psi$ to
\Ref{A} around the singularities $\l_k$ and $\l_l$, to be defined
explicitly below.

\section{Isomonodromic deformations on the Riemann sphere\\ and
Schlesinger transformations} 
 
\subsection{Schlesinger system on the Riemann sphere}
Consider the following ordinary linear differential equation for a
matrix-valued function $\Psi(\lambda)\in SL(2,\C)$:
\be
\f{d\Psi}{d\l}= A(\l)\,\Psi \equiv 
\sum_{j=1}^N\f{A_j}{\l-\l_j}\;\Psi\;,
\la{ls}\ee
where the residues $A_j\in \sl(2,\C)$ are independent of
$\l$. Regularity at $\l=\infty$ requires
\be
\sum_{j=1}^N A_j ~=~ 0\;,
\ee
and allows to further impose the initial condition
$\Psi(\l\!=\!\infty)=I$. 
The matrix $\Psi(\l)$ defined in this way lives on the
universal covering $X$ of $\C P^1\setminus\{\l_1,\dots,\l_N\}$. Its
asymptotical expansion near the singularities $\l_j$ is given by
\be
\Psi(\l)= G_j\Psi_j\cdot \,(\l-\l_j)^{T_j}\, C_j\;,
\la{asymp}\ee
with $G_j\,,C_j\,\in SL(2,\C)$ constant,
$\Psi_j=I+\cO(\l\!-\!\l_j)\in SL(2,\C)$ holomorphic around
$\l\!=\!\l_j$, and where $T_j$ is a traceless diagonal matrix with
eigenvalues $\pm t_j$. The residues $A_j$ of \Ref{ls} are encoded in
the local expansion as 
\be
A_j =G^{\phantom{1}}_j \,T^{\phantom{1}}_j\, G_j^{-1}\;.
\la{Aj}
\ee

Upon analytical continuation around $\l\!=\!\l_j$, the function
$\Psi(\lambda)$ in $\C P^1\setminus\{\l_1,\dots,\l_N\}$ changes by
right multiplication with some monodromy matrices $M_j$
\baa
\Psi(\l) &\to&  \Psi(\l)\,M_j\;,\la{Mj}\\[4pt]
M_j&=&C_j^{-1}\, e^{2\pi i T_j}\, C^{\phantom{1}}_j\;.\nn
\eaa
In the sequel we shall consider the generic case when none of $t_j$ is
integer or half-integer.
 
The assumption of independence of all monodromy matrices $M_i$ of the
positions of the singularities $\l_j$: $ \p M_i/\p\l_j=0$ is called
the isomonodromy condition; it implies the following dependence of
$\Psi(\l)$ on $\l_j$
\be
\f{\p\Psi}{\p\l_j}=-\f{A_j}{\l-\l_j}\;\Psi\;,
\la{ls1}\ee
as follows from \Ref{asymp} and normalization of $\Psi(\l)$ at
$\infty$. Compatibility of (\ref{ls}) and (\ref{ls1}) then is
equivalent to the classical Schlesinger system \c{Schl12}  
\be
\frac{\partial A_j}{\partial\l_i}=
\frac{[A_j,A_i]}{\l_j-\l_i}\;,\quad i\neq j\;,\qquad
\frac{\partial A_j}{\partial\l_j}=
-\sum_{i\neq j}
\frac{[A_j,A_i]}{\l_j-\l_i}\;.
\label{sch}
\ee
describing the dependence of the residues $A_j$ on the
$\l_i$. Obviously, the eigenvalues $t_j$ of the $A_j$ are integrals of
motion of the Schlesinger system. The functions $G_j$ accordingly have
the following dependance \cite{JiMiUe81}:
\be
\f{\p G_j}{\p \l_i} = \f{A_i G_j}{\l_i-\l_j}\;,\quad i\neq j\;,\qquad  
\f{\p G_j}{\p \l_j} = -\sum_{i\neq j}\f{A_i G_j}{\l_i-\l_j} \;,
\la{Geq}\ee
which obviously implies (\ref{sch}).

To introduce the notion of the $\tau$-function for the Schlesinger
system, one notes that \Ref{sch} is a multi-time Hamiltonian system
\c{JiMiMoSa80} with respect to the Poisson structure on the residues
$A_j$
\be
\left\{A_i^\A\,, A_j^\Ab\;\right\}= \d_{ij}\,\e^{\A\Ab\Ac}\,A_j^\Ac\;,
\la{Poisson}\ee
($\A,\Ab,\Ac$ denoting $\sl(2)$ algebra indices with the completely
antisymmetric structure constants $\e^{\A\Ab\Ac}$) and Hamiltonians
\be\la{Hs}
H_{i}~=~ \f{1}{4\pi\i}\oint_{\l_i} \tr A^2(\l)\,d\l ~=~
\ft12\,\sum_{j\neq i}\f{\tr A_i A_j}{\l_j\!-\!\l_i} \;.
\ee
Explicitly, \Ref{sch} takes the form
\be
\f{\p A_j}{\p \l_i}= \{H_i,  A_j\}\;,
\ee
and all the Hamiltonians $H_j$ Poisson-commute.

The $\tau$-function $\tau (\{\l_j\})$ of the Schlesinger system then
is defined as the generating functions of the Hamiltonians
\be
\f{\p\ln\tau}{\p\l_j} ~=~ H_j \;,
\la{tauHj}
\ee
where compatibility of these equations follows from \Ref{sch}. This
$\tau$-function is closely related to  the Fredholm determinant of a certain integral
operator associated to the Riemann-Hilbert problem (see
\cite{HarIts97} for details).

\subsection{Schlesinger transformations}

Schlesinger transformations are symmetry transformations of the
Schlesinger system (\ref{sch}) which map a given solution
$\left\{A_j(\{\l_i\})\right\}$ to another solution
$\left\{\hA_j(\{\l_i\})\right\}$ with the same number and positions of
poles $\l_j$ such that the related eigenvalues $t_j$ are shifted by
integer or half-integer values $t_j\to t_j\!+\!n_j/2\;,\;\; n_j\in
\Z$. The monodromy matrices $M_j$ hence remain invariant or change
sign under this transformation.  To be brief, we do not consider
Schlesinger transformations involving the point $\l=\infty$. Moreover,
we shall restrict ourselves to elementary Schlesinger transformations,
which change only two $t_j$'s, say, $t_k$ and $t_l$ for $k\neq l$ by
$\pm 1/2$. The transformed variables will be denoted by
$\hPsi,\,\hA_j,\, \tht_j$, etc. Without loss of generality we consider
the case
\be\la{tj}
\tht_j~=~ \left\{ \begin{array}{ll} t_j\!+\!\ft12\;\quad &
                               \mbox{for~} j=k,l \\  
                               t_j &\mbox{else} 
                               \end{array} \right. \;.
\ee
Our presentation here mainly follows \cite{Kita97}. For the transformed
function $\hPsi$ we make the ansatz
\be
\hPsi(\l) ~=~ F(\l)\,\Psi(\l) \;,
\la{Psih}\ee
with 
\be
F(\l)~=~ \sqrt{\f{\l-\l_k}{\l-\l_l}}\; S_+  +  
         \sqrt{\f{\l-\l_l}{\l-\l_k}}\; S_-  \;,
\la{Fsphere}
\ee
where the matrices $S_\pm$ do not depend on $\l$ and are uniquely
determined by \cite{Kita97}:
\be
S_\pm^2=S_\pm\;,\qquad S_++S_-=I\;,\qquad
S_+\,G_l^1= S_-\,G_k^1 =0\;.
\la{eqS}\ee
By $G_j^{\alpha}$ here we denote the $\alpha$-th column of the matrix
$G_j$ ($\alpha=1,2$). Combining the columns $G_k^1$ and $G_l^1$ into a
$2\times 2$ matrix 
\be
G=\left(G_k^1\,,  G_l^1\right), 
\la{Gkl}\ee
we can deduce from (\ref{eqS}) the following simple 
formula for $S_\pm$:
\be
S_\pm= G\, P_\pm\,  G^{-1} \;,
\la{SGkl}\ee
with projection matrices
\ben
P_+ =\left(\ba{cc} 1 & 0\\
                   0 & 0  \ea   \right)\;,\hskip1.3cm
P_- =\left(\ba{cc} 0 & 0\\
                   0 & 1  \ea   \right) \;.
\la{ppm}\een
It is easy to check using the local expansion of $\Psi$ at the
singularities $\l_j$ (\ref{asymp}) and the defining relations for
$S_\pm$ (\ref{SGkl}) that the transformed function $\hPsi$ at $\l_j$
has a local expansion of the form (\ref{asymp}) with the same matrices
$C_j$ and the desired transformation \Ref{tj} of the $t_j$. The
matrices $G_j$ change to new matrices $\hG_j$.  Thus, $\hPsi$
satisfies the system  
\be \la{lsh}
\f{\p\hPsi}{\p\l}= \sum_{j=1}^N\f{\hA_j}{\l\!-\!\l_j}\;\hPsi\;,
\qquad
\f{\p\hPsi}{\p\l_j}=-\f{\hA_j}{\l\!-\!\l_j}\;\hPsi\;,
\ee
where the functions $\hA_j(\{\l_i\})$ build a new solution of the
Schlesinger system (\ref{sch}).

On the level of the residues $A_j$, the form of the Schlesinger
transformation is not very transparent; however, it turns out that the
associated $\tau$-function transforms in a rather simple way.
Namely, for $\hPsi$ we find
\ben
\tr\,\hA^2 =  
\tr\,A^2 + 
2\,\tr\left[ F^{-1}\f{dF}{d \l}\; A\right]+
\tr\left[ F^{-1}\f{dF}{d\l}\right]^2 \;.
\een
The Hamiltonians $H_j$ for $j\neq k,l$ hence transform as 
\baa
\hH_j-H_j &=&\left(\f{1}{\l_j\!-\!\l_k}-\f{1}{\l_j\!-\!\l_l}\right)\,
\tr\left[A_j S_+\right] \\[4pt]
&=&\f{\tr\left[A_j G P_+ G^{-1}\right]}{\l_j-\l_k} + 
   \f{\tr\left[A_j G P_- G^{-1}\right]}{\l_j-\l_l} \non[4pt]
&\stackrel{\Ref{Geq}}{=}&
\tr\left[\;\f{\p G}{\p\l_j}\;G^{-1}\;\right] \non[4pt]
&=& \f{\p}{\p\l_j} \left\{\log\det G\right\} \;. \nn
\eaa
Therefore, the transformed $\tau$-function $\htau$ is given by $\htau
= f(\l_k,\l_l) \det G\,\cdot\, \tau$ with some function
$f(\l_k,\l_l)$ to be determined from the transformation of $H_k$,
$H_l$. In analogy to above we find that
\baa
\hH_k-H_k &=&
\sum_{j\neq k}\f{\tr \left[A_j S_+\right]}{\l_k\!-\!\l_j}
-\f{\tr \left[A_k S_+\right]}{\l_k\!-\!\l_l} -\f{1}{2(\l_k\!-\!\l_l)}  \\
&=& \f{\p}{\p\l_k} \log\det G -\f{1}{2(\l_k\!-\!\l_l)} \;,\nn
\eaa
and similarly for $H_l$. This yields the following formula describing
the action of elementary Schlesinger transformation \Ref{tj} on the
$\tau$-function:  
\be
\htau\left(\{\l_j\}\right) ~=~ \left\{(\l_k\!-\!\l_l)^{-1/2}\; 
\det G \right\}\,\cdot\,\tau\left(\{\l_j\}\right) \;.
\la{tauh}\ee
\bigskip

\begin{remark}\rm
The other elementary Schlesinger transformations like
\be\la{tjb}
\tht_j~=~ \left\{ \begin{array}{ll} t_j\!+\!\ft12\;\quad &
                               \mbox{for~} j=k \\[2pt]
                                    t_j\!-\!\ft12\;\quad &
                               \mbox{for~} j=l \\ 
                               t_j &\mbox{else} 
                               \end{array} \right. \;,
\ee
etc., may be obtained in a similar way by building the matrix $G$
from $G^1_k$ and $G^2_l$ instead of \Ref{Gkl}, etc..  Moreover, all
such transformations with different $k$ and $l$ may be superposed to
get the general Schlesinger transformation which simultaneously shifts
an arbitrary number of the $t_j$ by some integer or half-integer
constants.  These general transformations were in detail studied in
\cite{JiMiUe81,JimMiw81b,JimMiw81c}. In their framework, $\det G$
from \Ref{tauh} is introduced as particular matrix element
$\left(G_k^{-1}G_l^{\vphantom{-1}}\,\right)_{12}$ (an identity which
holds for $2\times2$ matrices).  
\end{remark}

\begin{remark}\rm
Carefully comparing \Ref{tauh} to the result of \cite{JimMiw81b}, we
find an additional factor of $(\l_k\!-\!\l_l)^{-1/2}$. This is due to
the fact, that we treat the $\sl(2,\C)$ case rather than the
$\gl(2,\C)$ case which is done in \cite{JimMiw81b}. Indeed, this
amounts to a simple renormalization of the $\Psi$-function by e.g.
$\sqrt\frac{\l-\l_k}{\l-\l_l}$ after the Schlesinger transformation
\Ref{tj}, which generates precisely this additional factor.

\end{remark}

\section{Schlesinger transformations for isomonodromic deformations on
the torus} 
In this section, we generalize the construction of Schlesinger
transformations described above to the case of genus one Riemann
surfaces. To this end we first review some basic elliptic functions
and subsequently the isomonodromic deformations on the torus in the
setting of \c{Taka98}.

\subsection{Some elliptic functions}
 
The elliptic theta-function with characteristic $[p,q]$ ($p,q\in \C$)
on a torus $E$ is defined by the series
\be
\th[p,q](\l|\mu)=\sum_{m\in \Z} e^{\pi \i\mu (m+p)^2 +2\pi \i 
(m+p)(\l+q)} \;. 
\ee
Let us introduce on the torus $E$ the standard Jacobi theta-functions:
\baa
\th_1(\l)&\equiv& - \th\left[\ft{1}{2},\ft{1}{2}\right](\lambda|\mu) \;,
\\
\th_2(\l)&\equiv& \th\left[\ft{1}{2},0\right](\lambda|\mu) \;,
\non
\th_3(\l)&\equiv&\th(\l)\equiv  \th[0,0](\lambda|\mu) \;,
\non
\th_4(\l)&\equiv& \th\left[0,\ft{1}{2}\right](\lambda|\mu)\;,
\nn
\eaa
and corresponding theta-constants
\ben
\th_j\equiv \th_j(0)\;,\hskip0.8cm j=2,3,4\;.
\een
We define the following three combinations of Jacobi theta-functions:
\be
w_1(\l)= \pi\th_2\th_3 \f{\th_4(\l)}{\th_1(\l)}\;,\quad
w_2(\l)= \pi\th_2\th_4 \f{\th_3(\l)}{\th_1(\l)}\;,\quad
w_3(\l)= \pi\th_3\th_4 \f{\th_2(\l)}{\th_1(\l)}\;.
\la{wj}\ee
All these functions have simple poles at $\l=0$ with residue
1. Moreover, they possess the following periodicity properties: 
\baa
w_1(\l+1) = - w_1(\l) \quad&&\qquad w_1(\l+\mu) = \phantom{-}w_1(\l)\;, 
\la{perw} \\
w_2(\l+1) = - w_2(\l) \quad&&\qquad w_2(\l+\mu) = - w_2(\l)\;, \non
w_3(\l+1) = \phantom{-}  w_3(\l) \quad&&\qquad w_3(\l+\mu) = - w_3(\l)\;.
\nn
\eaa
In the sequel we shall also need the following functions $\Zf_\A$:
\be
\Zf_1= \f{w_1}{2\pi i}\f{\th_4'(\l)}{\th_4(\l)} \;,\quad
\Zf_2= \f{w_2}{2\pi i}\f{\th_3'(\l)}{\th_3(\l)}\;,\quad
\Zf_3= \f{w_3}{2\pi i}\f{\th_2'(\l)}{\th_2(\l)}\;.
\la{Z}\ee
with the following periodicity properties:
\baa
\Zf_1 (\l+1) = - \Zf_1 (\l) \quad&&\qquad 
\Zf_1 (\l+\mu) = \phantom{-}\Zf_1 (\l) - w_1\;, \la{perZ} \\ 
\Zf_2 (\l+1) =  - \Zf_2 (\l) \quad&&\qquad 
\Zf_2 (\l+\mu) =  - \Zf_2 (\l) + w_2\;, \non
\Zf_3 (\l+1) = \phantom{-} \Zf_3 (\l) \quad&&\qquad 
\Zf_3 (\l+\mu) =  - \Zf_3 (\l) + w_3\;. \nn
\eaa
It is easy to verify the identity
\be
\f{d w_\A}{d\mu}(\l)= \f{d \Zf_\A}{d\l}(\l)\;.
\la{cross}\ee
Let us check this for example, for $\mbox{{\sc a}}=1$. Both sides of
(\ref{cross}) are holomorphic in $E$. Moreover, from the periodicity
properties of $w_1$ and $\Zf_1$ we have: 
\ben
\f{d\Zf_1}{d\l} (\l+1) = - \f{d\Zf_1}{d\l} (\l)\hskip1.0cm
\f{d\Zf_1}{d\l} (\l+\mu) = \f{d\Zf_1}{d\l} (\l) - \f{dw_1}{d\l}(\l) \;.
\een
Therefore, the difference ${dw_1}/{d\mu}- {d\Zf_1}/{d\l}$ is
holomorphic in $E$, invariant with respect to the $\mu$-shifts of $\l$
and changes sign with respect to unit shifts of $\l$. It hence
vanishes and  (\ref{cross}) is fullfilled. 

Let us list some further useful properties of the functions $w_\A$ and
$\Zf_\A$:

\begin{itemize}\rm

\item
The functions $w_\A$ may be represented as ratios of Jacobi's elliptic
functions as follows: 
\be
w_1(\l)= \f{1}{{\rm sn}(\l)\;}\;,\quad 
w_2(\l)= \f{{\rm dn}(\l)}{{\rm sn}(\l)}\;,\quad 
w_3(\l)= \f{{\rm cn}(\l)}{{\rm sn}(\l)} \;.
\la{wsn}\ee

\item
The functions $w_\A$ satisfy the following differential equation:
\be
\f{dw_1(\l)}{d\l}= -  w_2(\l) w_3(\l) \;,
\la{wprime}\ee
and cyclic permutations thereof. This relation may be easily proved
comparing the behaviour at $\l=0$ and the twist properties of both sides
of (\ref{wprime}). 

\item
The functions $w_\A$ satisfy the following summation theorem which
easily follows from the summation theorem for Jacobi functions:
\be
w_1(\l\!+\!\l')- w_1(\l\!-\!\l') ~=~
\f{2w_1(\l')w_2(\l)w_3(\l)}{w_1^2(\l)- w_1^2(\l')} \;,
\la{sumtheo}\ee
and cyclic permutations thereof. At $\l=\l'$ this relation
implies   
\be\la{div}
2w_1(2\l) ~=~ \f{\p}{\p\l}\,\ln\left(\f{w_1}{w_2w_3} \right) \;.
\ee

\item
For any values of $\mbox{{\sc a, b}}$, the difference of squares
$w_\A^2(\l) - w_\Ab^2(\l)$ is independent of $\l$, as follows from its
single-valuedness and holomorphy on $E$. From the well-known relations
between the squares of Jacobi elliptic functions we find more
precisely that $w_1^2(\l)\!-\!w_2^2(\l)=x$;\,
$w_1^2(\l)\!-\!w_3^2(\l)=1$ where $x=x(\mu)$ is the module parameter
of the torus $E$ which arises from realizing the torus as two-sheet
covering of the complex plane with branch points $0,1,x,\infty$.

\item
The previous property in particular implies that the expression
$w_\A^2(\l)- w_\A^2(\l')$ does not depend on $\mbox{{\sc a}}$ for any
values of $\l$ and $\l'$.

\item
We have the following formula for integration of the product of two
functions $w_\A$ along the basic $a$-cycle of the torus $E$ 
\be
\oint_a w_\A(\l\!-\!\l_1)w_\A(\l\!-\!\l_2) \,d\l ~=~ 
2\pi \i\,\Zf_\A (\l_1\!-\!\l_2) \;.
\la{intww}\ee
This formula can be verified by checking the analyticity and periodicity 
properties of both sides in, say, the $\l_1$-plane.

\end{itemize}

\subsection{Isomonodromic deformations on the torus}

Consider the elliptic curve $E$ with periods $1$ and $\mu$ together
with the canonical basis of cycles $(a,b)$. A (naive) straightforward
generalization of the idea of isomonodromic deformations from the
complex plane to the torus $E$ runs into difficulties related to the
absence of meromorphic functions on the torus with just one simple
pole. An independent variation of the simple poles of a meromorphic
connection $A$ on the torus preserving the monodromies around the
singularities and basic cycles is impossible for the following simple
reason. Existence of such a deformation would imply a version of
(\ref{ls1}) with the function $\f{A_j}{\l-\l_j}$ on the r.h.s.~being
substituted by a meromorphic function with only one simple pole on the
torus, which gives rise to the contradiction.  Therefore, one of the
underlying assumptions has to be relaxed.

E.g.~one may consider the case where not all the poles of the
connection $A$ are varied independently. Another possibility is the
assumption that some of the poles of $A$ are of order higher then one
\cite{Okam79}. A third alternative which we shall consider here, is to
relax the condition of single-valuedness of the connection $A$ on $E$
and assume that $A$ has ``twists'' with respect to analytical
continuation along the basic cycles $a$ and $b$, i.e.
\ben
A(\l+1) = Q A(\l) Q^{-1}\;,\qquad
A(\l+\mu) = R A(\l) R^{-1} \;,
\een
where the matrices $Q,\,R$ do not depend on $\l$. By a gauge
transformation of the form $A\to S A S^{-1} + dS S^{-1}$ with $S$
holomorphic but possibly multi-valued, one may bring the connection
into a form where $Q=I$ and $R=e^{\ka \sigma_3}$,  where $\sigma_\A$
denote the Pauli matrices: 
\ben
\sigma_1 =\left(\ba{cc} 0 & 1\\
                       1 & 0  \ea   \right)\;,\qquad
\sigma_2 =\left(\ba{cc} 0  & \i\\
                 -\i & 0       \ea\right)\;,\qquad
\sigma_3 =\left(\ba{cc} 1  & 0\\
                       0 & -1      \ea\right)\;.
\een
The equations of isomonodromic deformations with this choice of the
twist were considered in \cite{KorSam97} where the multi-valuedness of
$A$ had a natural origin in the holomorphic gauge fixing of
Chern-Simons theory on the punctured torus. The resulting equations
however are rather complicated in comparison with the Schlesinger
system on the sphere.  This is due to the fact that the twist $\ka$
itself becomes a dynamical variable -- i.e.~changes under
isomonodromic deformations -- and in generic situation has a highly
non-trivial $\l_j$-dependence.  Therefore, instead of being bilinear
with respect to the dynamical variables, this Schlesinger system on
the torus becomes highly transcendental.
  
An alternative form of the elliptic Schlesinger system was proposed by
Takasaki \cite{Taka98} who considered the restriction $Q=\sigma_3$,
$R=\sigma_1$, related to the classical limit of Etingof's elliptic
version of the Knizhnik-Zamolodchikov-Bernard system on the torus
\c{KniZam84,Bern88a,Etin94}. This choice of fixing the twists turns
out to be compatible with the isomonodromic deformations equations,
therefore essentially simplifying the dynamics as compared to
\cite{KorSam97}.  It results into studying isomonodromic deformations
of the system
\baa
\f{d\Psi}{d\l}&=& A(\l)\, \Psi\;, \la{lstor} \\
A(\l)&\equiv&\sum_{j=1}^N \sum_{\A=1}^3 A^{\A}_j \,w_\A(\l\!-\!\l_j) 
\;, \nn
\eaa
with $\l\in \C$ and functions $w_\A$ from \Ref{wj}. The connection
$A(\l)$ obviously has only simple poles on $E$ and the following twist
properties, cf.~\Ref{perw}  
\be
A(\l+1)= \sigma_3 \,A(\l)\, \sigma_3 \;,\qquad
A(\l+\mu)= \sigma_1\, A(\l) \,\sigma_1\;.
\la{Taktwist}\ee
Since the residues of all $w_\A$ at $\l=0$ coincide, the residue of
$A(\l)$ at $\l_j$ is
\ben
A_j \equiv \sum_{\A} A^{\A}_j \, \sigma_\A\;.
\een                                                                           
As in the case of the Riemann sphere, the function $\Psi$ has regular
singularities at $\l=\l_j$ with the same local properties
\Ref{asymp}--\Ref{Mj}. The twist properties of $\Psi$ take the
form 
\be
\Psi(\l+1) = \sigma_3  \Psi(\l) M_a \hskip1.0cm
\Psi(\l+\mu) = \sigma_1  \Psi(\l) M_b \;,
\la{perPsi}\ee
with monodromy matrices $M_a$, $M_b$ along the basic cycles of the
torus.  Moreover, as in the case of Riemann sphere, $\Psi(\l)$  
has monodromies $M_j$ around the singularities $\l_j$.

The isomonodromy condition on the torus implies that all monodromies
$M_j$, $M_a$ and $M_b$ are independent of the positions of
singularities $\l_j$ and the module $\mu$ of the torus. As on the
Riemann sphere this implies that the function
${\p\Psi}/{\p\l_j}\,\Psi^{-1}$ has the only simple pole at
$\l\!=\!\l_j$ with residue $-A_j$. In addition, it has the following
twist properties
\baa
\f{\p\Psi}{\p{\l_j}}\,\Psi^{-1} (\l+1)&=& 
\sigma_3\,\f{\p\Psi}{\p\l_j}\,\Psi^{-1} (\l)\,\sigma_3 \;,
\non
\f{\p\Psi}{\p\l_j}\,\Psi^{-1} (\l+\mu)&=& 
\sigma_1\,\f{\p\Psi}{\p\l_j}\,\Psi^{-1} (\l)\,\sigma_1 \;.
\nn
\eaa
Therefore,
\be
\f{\p\Psi}{\p\l_j} = - \sum_{\A=1}^3 A^{\A}_j\, w_\A(\l-\l_j) \sigma_\A\, 
\Psi\;.
\la{Psigjtor}\ee
To derive the equation with respect to module $\mu$ we observe that
$\p\Psi/\p\mu\,\Psi^{-1}$ is holomorphic at $\l\!=\!\l_j$ (but not at
$\l=\l_j\!+\!\mu$\,) and has twist properties  
\baa
\f{\p\Psi}{\p\mu}\,\Psi^{-1} (\l+1) &=&
\sigma_3\, \f{\p\Psi}{\p\mu}\,\Psi^{-1} (\l)\,\sigma_3 \;,
\non
\f{\p\Psi}{\p\mu}\,\Psi^{-1} (\l+\mu) &=&
\sigma_1\,\left(\f{\p\Psi}{\p\mu}\,\Psi^{-1} (\l)- 
\f{\p\Psi}{\p\l}\,\Psi^{-1} (\l)\right)\,\sigma_1 \;.
\nn
\eaa
Taking into account the periodicity properties of the functions
$\Zf_\A$ (\ref{perZ}), this hence implies
\be
\f{\p\Psi}{\p\mu}= \sum_{j=1}^N \sum_{\A=1}^3 A^{\A}_j\, 
\Zf_\A \sigma_\A\;\Psi \;.
\la{Psitau}\ee
The compatibility conditions of the equations (\ref{lstor}),
(\ref{Psigjtor}) and  (\ref{Psitau}) then yield the $\l_i$ and $\mu$
dependence of the residues $A_j$. The result is summarized in the
following 
\begin{theorem}
{\rm \c{Taka98}} Isomonodromic deformations of the system
(\ref{lstor}) are described the the following elliptic version of the
Schlesinger system: 
\baa
\f{d A_j}{d \l_i}&=& \left[\,A_j\,, \sum_{\A=1}^3 A_i^\A\, 
w_\A(\l_j\!-\!\l_i)\,\sigma_\A\,\right]\;,\qquad i\neq j \;,\la{Se1}\\[2pt]
\f{d A_j}{d \l_j}&=&-\sum_{i\not=j}
\left[\,A_j\,, \sum_{\A=1}^3 A_i^\A\, w_\A (\l_j\!-\!\l_i)\,
\sigma_\A\,\right] \;,\non[2pt]
\f{d A_j}{d \mu}&=& 
-\sum_{i=1}^N \left[\,A_j\,, \sum_{\A=1}^3 A_i^\A\, 
\Zf_\A (\l_j\!-\!\l_i)\,\sigma_\A\,\right] \;.
\nn
\eaa
\end{theorem} 
\qed

\medskip

The corresponding equations for the matrices $G_j$ from (\ref{asymp})
take a form analogous to the equations (\ref{Geq}) on the Riemann
sphere: 
\be
\f{\p G_j}{\p\l_i}=\sum_\A A_i^\A\, w_\A(\l_i\!-\!\l_j)\,\s_\A\; G_j\;,\qquad
\f{\p G_j}{\p\l_j}= -\sum_{i=1}^N \sum_\A  
A_i^\A\, w_\A(\l_i\!-\!\l_j)\,\s_\A\; G_j \;.
\la{Gell}\ee
The system (\ref{Se1}) admits a multi-time Hamiltonian formulation
with respect to the Poisson structure \Ref{Poisson} on the residues
\ben
\{A_i^\A, A_j^\Ab\}= \d_{ij}\,\e^{\A\Ab\Ac}\,A_j^\Ac \;.
\een
This is summarized as
\begin{theorem}
The elliptic Schlesinger system \Ref{Se1} is a multi-time Hamiltonian
system with respect to the Poisson bracket 
\be
\Big\{\stackrel1{A}\!(\l), \stackrel2{A}\!(\l')\Big\} = 
\left[\;r(\l-\l')\,, \stackrel1{A}\!(\l)\;+
\stackrel2{A}\!(\l')\right] \;,
\la{rmatr}\ee
with the elliptic classical $r$-matrix $r$ given by 
\ben
r(\l) = \sum_\A w_\A(\l)\, \s_\A\otimes \s_\A \;.
\een
The Hamiltonians describing deformation with respect to the variables
$\l_i$ and to the module $\mu$ of the torus are respectively given by
\baa
H_i &=& \f{1}{4\pi i}\oint_{\l_i}\tr A^2(\l)d\l ~=~
\sum_{j\neq i} \sum_{a} A_j^\A A_i^\A \, w_\A(\l_j-\l_i) \;,
\la{Hjcon}\\[6pt]
H_\mu &=& -\f{1}{2\pi i}\oint_a \tr A^2(\l)d\l ~=~
- \sum_{i,j} \sum_{a}   A_i^\A A_j^\A \, {\cal{Z}}_\A(\l_i-\l_j) 
\;, \la{Hmucon}
\eaa
and mutually Poisson commute.
\end{theorem}
This follows from straight-forward calculation. The representation of
$H_\mu$ as contour integral along the basic $a$-cycle in
(\ref{Hmucon}) may be derived using the relations (\ref{intww}). The
fact, that all Hamiltonians Poisson-commute is a direct consequence of
\be
\left\{\,\tr A^2(\l)\,,\tr A^2(\l')\,\right\} ~=~ 0 \;,
\ee
which in turn follows immediately from \Ref{rmatr}. \qed

\medskip

Now we are in position to define the $\tau$-function of the elliptic
Schlesinger system \Ref{Se1} as generating function
$\tau\left(\{\l_j\},\mu\right)$ of the Hamiltonians   
\be
\f{\p\ln\tau}{\p\l_j} = H_j\;,\qquad
\f{\p\ln\tau}{\p\mu} = H_\mu \;,
\la{tauell}\ee 
whereby it is uniquely determined up to an arbitrary
$(\mu,\{\g_j\})$-independent multiplicative constant. As ususal,
consistency of this definition is a corollary of the elliptic
Schlesinger system.

\subsection{Schlesinger transformations for elliptic isomonodromy
deformations} 

The natural generalization of the notion of Schlesinger
transformations on the Riemann sphere as introduced above to the
elliptic case is the following. Starting from any solution of the
elliptic Schlesinger system (\ref{Se1}) with associated function
$\Psi$ satisfying (\ref{lstor}) and (\ref{perPsi}) we construct a new
solution $\hA_j$, $\hPsi$ with eigenvalues $\tht_j$ which differ
from the $t_j$ by integer or half-integer values. In particular, we
will consider the elliptic analog of the elementary Schlesinger
transformation \Ref{tj} on the Riemann sphere.  The following
construction is inspired by the papers \cite{BiBoIt84}, \cite{DJKM83}.

As a natural elliptic analog of the function $F(\l)$ from
(\ref{Fsphere}) we shall choose the following ansatz
\baa
F(\l) &=& \f{f(\l)}{\sqrt{\det f(\l)}} \;,\la{Fell}\\
f(\l) &=& \f{1}{2} +\sum_{\A=1}^3 J_\A\, 
w_\A\!\left(\l-\ft12(\l_k\!+\!\l_l)\right)\,\s_\A \;,\nn
\eaa
where the functions $J_\A(\l_j,\mu)$ depend on $G_k$ and $G_l$ and
will be defined below. We formulate the result of this section in the
following  
\begin{theorem}\la{dresstor}

Let the functions $\left\{A_j(\{\l_i\})\right\}$ satisfy the elliptic
Schlesinger system (\ref{Se1}) with twist properties (\ref{Taktwist})
and let the function $\Psi$ satisfy the associated linear system
(\ref{lstor}). For two arbitrary non-coinciding poles $\l_k$ and
$\l_l$, define the new function
\be
\hPsi(\l)~\equiv~ F(\l)\,\Psi(\l) \;,
\la{Psihell}\ee
with $F(\l)$ from formula (\ref{Fell}) with $\l$-independent
coefficients $J_\A$ defined by  
\be
\sum_\A J_\A\, w_\A\!\left(\ft12\,(\l_k\!-\!\l_l)\right)\,\s_\A ~\equiv~  
- \ft{1}{2}\, G \,\s_3\,  G^{-1} \;,
\la{SGell}\ee
where as above we denote by $G$ the matrix \Ref{Gkl} containing
the first columns of the matrices $G_k$ and $G_l$. 

The function $\hPsi(\l)$ then satisfies the equations (\ref{lstor}),
(\ref{Psigjtor}), (\ref{Psitau}) and the twist conditions
(\ref{perPsi}) with the transformed functions 
\be
\hA_j
\left(\{\l_i\}\right)~\equiv~
\res_{\l=\l_j}\left\{\f{d\hPsi}{d\l}\hPsi^{-1}\right\} \;. 
\la{hAjell}\ee
In turn, the functions $\hA_j$ satisfy the elliptic Schlesinger system
(\ref{Se1}). For the eigenvalues $t_j$ we have
\ben
\tht_j~=~ \left\{ \begin{array}{ll} t_j\!+\!\ft12\;\quad &
                               \mbox{for~} j=k,l \\  
                               t_j &\mbox{else} 
                               \end{array} \right. \;.
\een
The monodromy matrices $\hM_j$, $\hM_a$ and $\hM_b$ of
the function $\hPsi$ coincide with the monodromies of $\Psi$, except
for $\hM_k=-M_k$ and  $\hM_l=-M_l$. 
\end{theorem}
\medskip

{\it Proof.} The proof consists of two parts. The first part is to
check that locally in the neighbourhoods of the singularities $\l_j$
the situation looks exactly like the situation on the Riemann
sphere. The second (global) part is to check that no new singularities
arise apart from the $\l_j$ and that the new function $\hPsi$
satisfies the required twist conditions (\ref{perPsi}).

The proper local behaviour of function $\hPsi$ is ensured by the
relations  
\be\la{Sell}
S_\pm^2=S_\pm\;,\qquad S_++S_-=I\;,\qquad
S_+\,G_l^1= S_-\,G_k^1 =0\;;
\ee
for
\ben
S_\pm ~\equiv~ \f{1}{2}\mp \sum_\A J_\A\; 
w_\A\!\left(\ft12\,(\l_k\!-\!\l_l)\right)\, \s_\A ~=~
G \,P_\pm\,  G^{-1} \;,
\een
which in complete analogy to (\ref{eqS}) describe annihilation of the
vectors $G_k^1$ and $G_l^1$ by the matrices $f(\l_k)$ and $f(\l_l)$,
respectively. Obviously, equations \Ref{Sell} are a consequence of
\Ref{SGell}. Similarly to the case of the sphere, it is then easy to
verify that (\ref{Sell}) provide the required asymptotical expansions
(\ref{asymp}) for the function $\hPsi$ with parameters $\hG_j$, $C_j$
and $\tht_j$. 

Concerning the global behavior of $\hPsi$ we note that the prefactor
$(\det f(\l))^{-1/2}$ in (\ref{Fell}) provides the condition
$\det\hPsi =1$ and kills the simple pole of $f(\l)$ at
$\l=(\l_k\!+\!\l_l)/2$. Therefore, the only singularities of $F(\l)$
on $E$ are the zeros of $\det f(\l)$. Since $\det f(\l)$ has only one
pole -- this is the second order pole at $\l=(\l_k\!+\!\l_l)/2$ -- it
must have also two zeros on $E$ whose sum according to Abel's theorem
equals $\l_k+\l_l$.  According to (\ref{Sell}) these are precisely
$\l_k$ and $\l_l$.  It remains to check that $\hPsi$ satisfies
conditions (\ref{perPsi}) with the same matrices $M_a$ and $M_b$.
This follows from the twist properties
\ben
f(\l+1)~=~ \s_3 \,f(\l)\, \s_3\;,\qquad 
f(\l+\mu)~=~ \s_1\, f(\l) \,\s_1\;,
\een
which in turn is a consequence of (\ref{Fell}) and the periodicity
properties (\ref{perw}) of the functions $w_j(\l)$. 

\qed

\medskip

\mathversion{bold}
\section{The action of elliptic Schlesinger transformations\\ on the
$\tau$-function}
\mathversion{normal}

In this section we shall present and prove the elliptic analog of
formula (\ref{tauh}) describing the transformation of the
$\tau$-function under the action of elliptic Schlesinger
transformations. 

\begin{theorem}
The $\tau$-function $\htau$ corresponding to the Schlesinger-transformed
solution $\hA_j$ (\ref{hAjell}) of the elliptic Schlesinger system is
related to the $\tau$-function corresponding to the solution $A_j$ as
follows 
\be\la{tauhtau}
\htau\left(\{\l_j\},\mu\right)~=~ 
\left\{\Big[w_1 w_2 w_3\!\left(\f{\l_k\!-\!\l_l}{2}\right)\; 
\Big]^{1/2}\;\det\,\Big[ G J^{1/2}\Big]\:\right\}\,
\cdot\, \tau\left(\{\l_j\},\mu\right) \;,
\ee
where $G$ is the matrix \Ref{Gkl} containing the first columns of
the matrices $G_k$, $G_l$; $J$ is the matrix
$$
J\equiv \sum_{\A=1}^3 J_A \sigma_A
$$
and the functions $J_\A$ are defined in terms of  $G$ via (\ref{SGell}).  
\end{theorem}
\medskip

{\it Proof.}
We start noting that
\be
\sum_\A \left[J_\A\, w_\A\!\left(\ft12(\l_k\!-\!\l_l)\right)\right]^2 ~=~
\f{1}{4} \;,
\la{detSG}
\ee
as follows from taking the determinant of (\ref{SGell}). In
particular, this shows that upon replacing $w_\A(\l)\rightarrow 1/\l$,
formula \Ref{tauhtau} indeed reproduces the result for the Riemann
sphere \Ref{tauh}. 

The proof of \Ref{tauhtau} according to the definition of the
$\tau$-function (\ref{tauell}) now consists of three parts; it has to
be checked that
\begin{subeqnarray}\la{HHH}
\hH_j-H_j &~=~& \f{\p}{\p\l_j}\;\ln\left\{\Big[\sum_{\A=1}^3 J_A^2
\Big]^{1/2}\;\det G\:\right\}
\quad\mbox{for}\quad j\not=k,l\;,\\[4pt]
\hH_k-H_k &~=~& \f{\p}{\p\l_k}\;
\ln\left\{\Big[w_1 w_2 w_3\!\left(\ft12\,(\l_k\!-\!\l_l)\right)\; 
\sum_{\A=1}^3 J_A^2
\Big]^{1/2}\;\det G\:\right\}\;,\\[4pt]
\hH_\mu-H_\mu &~=~& \f{\p}{\p\mu}\;
\ln\left\{\Big[w_1 w_2 w_3\!\left(\ft12\,(\l_k\!-\!\l_l)\right)\; 
\sum_{\A=1}^3 J_A^2
\Big]^{1/2}\;\det G\:\right\}\;. 
\end{subeqnarray}
To obtain the l.h.s.~of these equations we make use of the
representation of the Hamiltonians as contour integrals (\ref{Hjcon}),
(\ref{Hmucon}) and compute the difference
\be
\tr\hA^2 - \tr A^2 ~=~  
2\,\tr\left[ F^{-1}\,\f{dF}{d \l}\, A\;\right]+
\tr\left[ F^{-1}\,\f{dF}{d\l}\right]^2 \;.
\la{trAh}\ee
It is
\baa
F^{-1}\,\f{dF}{d\l}&=& -\f{2}{1-4\sum_\A (J_\A w_\A)^2}\;\times 
\non[4pt]
&&\qquad\quad \times
\left(\sum_{a} J_\A w_\A' \s_\A - \i \sum_{\A,\Ab,\Ac}
\e^{\A\Ab\Ac}\, J_\Ab J_\Ac \,w_\A\, (w_\Ab^2-w_\Ac^2)\,\s_\A \right)
\left(\l-\ft12(\l_k+\l_l)\right) \;,
\nn
\eaa
where all the elliptic functions $w_\A$ on the r.h.s.~are taken at the
argument $\l-\ft12(\l_k+\l_l)$. Making use of \Ref{wprime},
\Ref{sumtheo}, \Ref{detSG} and the fact that the combination
$w_\A^2(\l)\!-\!w_\A^2(\l')$ does not depend on $\mbox{{\sc a}}\,$, this
expression simplifies to
\baa
F^{-1}\f{dF}{d\l}&=& -\f1{4\sum_\A (J_A^2)}\;\sum_\A J_\A\,
\f{w_\A(\l\!-\!\l_k)-w_\A(\l\!-\!\l_l)}
{w_\A\!\left(\ft12(\l_k\!-\!\l_l)\right)}\;\s_\A
\la{FFl}\\[4pt]
&&{}-\f{\i}{4\sum_\A (J_A^2)}\, \sum_{\A,\Ab,\Ac}
\e^{\A\Ab\Ac}\, J_\Ab J_\Ac \,(w_\Ab^2-w_\Ac^2)\,
\f{w_\A(\l\!-\!\l_k)+w_\A(\l\!-\!\l_l)}
{w_\Ab w_\Ac\!\left(\ft12(\l_k\!-\!\l_l)\right)} \;\s_\A \;.\nn
\eaa
The argument in the combinations $w_\Ab^2-w_\Ac^2$ has been skipped
since they are $\l$-independent. 
\bigskip

Let us start by proving (\ref{HHH}a). According to \Ref{trAh} and
\Ref{FFl} we find that  
\baa
\hH_j-H_j &=& 
-\f1{2\sum_\A (J_A^2)}\;\sum_\A J_\A A_j^\A\;
\f{w_\A(\l_j\!-\!\l_k)-w_\A(\l_j\!-\!\l_l)}
{w_\A\!\left(\ft12(\l_k\!-\!\l_l)\right)}
\la{HjhHj}\\[4pt]
&&{}-\f{\i}{2\sum_\A (J_A^2)}\, \sum_{\A,\Ab,\Ac}
\e^{\A\Ab\Ac}\, A_j^\A\,J_\Ab J_\Ac \,(w_\Ab^2-w_\Ac^2)\,
\f{w_\A(\l_j\!-\!\l_k)+w_\A(\l_j\!-\!\l_l)}
{w_\Ab w_\Ac\!\left(\ft12(\l_k\!-\!\l_l)\right)}  \;.\nn
\eaa
This should be compared to the r.h.s~of (\ref{HHH}a). From (\ref{Gell})
we find that 
\ben
\f{\p G}{\p\l_j} = \sum_\A A_j^\A\s_\A
\left\{w_\A(\l_j\!-\!\l_k)\,GP_+ + 
w_\A(\l_j-\l_l)\,GP_-\right\}\;.
\een
Thus,
\baa
\f{\p}{\p\l_j} \left\{\log\det G\right\} &=&
\tr\left[\;\f{\p G}{\p\l_j}\;G^{-1}\;\right]
 \la{ljdetG}\\[4pt]
&=& -2\,\sum_\A A_j^\A\,J_\A\,w_\A\!\left(\ft12(\l_k\!-\!\l_l)\right)\,
 \left(w_\A(\l_j\!-\!\l_k)-
w_\A(\l_j-\l_l)\right) \;. \nn
\eaa
It is slightly more complicated to calculate the first term in the
r.h.s.~of (\ref{HHH}a). Equations (\ref{SGell}) and (\ref{Gell}) imply
that  
\baa
2\sum_\A\f{\p J_\A}{\p\l_j} 
w_\A\!\left(\ft12(\l_k\!-\!\l_l)\right) \s_\A &=& 
-\f{\p}{\p\l_j} 
\Big\{ G^{\vphantom{-1}}\,\s_3\, G^{-1}\Big\} \la{inst}\\[4pt]  
&=& 
-\sum_\A A_j^\A\,\left(w_\A(\l_j\!-\!\l_k)\,S_- -
 w_\A(\l_j\!-\!\l_l)\,S_+\right)\,\s_\A \non
&&
{} + G^{\vphantom{-1}}\,\s_3\,G^{-1}\;\sum_\A
A_j^\A\,\left(w_\A(\l_j\!-\!\l_k)\,S_- -
w_\A(\l_j\!-\!\l_l)\,S_+ \right)\,\s_\A\;. \nn
\eaa
Reexpressing $G\,\s_3\, G^{-1}$ and $S_\pm$
in terms of $J_\A$ according to (\ref{SGell}) and \Ref{Sell}, we get
after some calculation 
\baa
\f{\p}{\p\l_j}\sum_\A J_A^2 &=&
-\sum_\A J_\A A_j^\A\;
\f{w_\A(\l_j\!-\!\l_k)-w_\A(\l_j\!-\!\l_l)}
{w_\A\!\left(\ft12(\l_k\!-\!\l_l)\right)}\;
\left(1-4w_\A^2\!\left(\ft12(\l_k\!-\!\l_l)\right)\,\sum_bJ^2_\Ab \right)
\non[4pt]
&&{}-\i \sum_{\A,\Ab,\Ac}
\e^{\A\Ab\Ac}\, A_j^\A\,J_\Ab J_\Ac \,(w_\Ab^2-w_\Ac^2)\,
\f{w_\A(\l_j\!-\!\l_k)+w_\A(\l_j\!-\!\l_l)}
{w_\Ab w_\Ac\!\left(\ft12(\l_k\!-\!\l_l)\right)}  \;,\nn
\eaa
such that combining this with \Ref{HjhHj} and \Ref{ljdetG} we indeed
recover (\ref{HHH}a). This determines the function $\htau$ already up to
some factor depending on $\l_k$, $\l_l$ and $\mu$ in accordance with
formula (\ref{tauhtau}). 
\bigskip

We turn to proving (\ref{HHH}b). According to \Ref{trAh}, also the
residues of $F^{-1}\,dF/d\l$ at $\l_k$ enter the variation of
$H_k$. Expanding \Ref{FFl} we find 
\baa
F^{-1}\f{dF}{d\l}&=& -\f1{4\,(\l-\l_k)\,\sum_\A (J_A^2)}\;
\left\{\sum_\A 
\f{J_\A\,\s_\A}
{w_\A\!\left(\ft12(\l_k\!-\!\l_l)\right)}
+ \sum_{\A,\Ab,\Ac}
\f{\i\e^{\A\Ab\Ac}\, (w_\Ab^2\!-\!w_\Ac^2)\,J_\Ab J_\Ac\,\s_\A} 
{w_\Ab w_\Ac\!\left(\ft12(\l_k\!-\!\l_l)\right)} \right\}  \non[6pt]
&&{}+\f1{4\sum_\A (J_A^2)}\;\sum_\A J_\A\,
\f{w_\A(\l_k\!-\!\l_l)}
{w_\A\!\left(\ft12(\l_k\!-\!\l_l)\right)}\;\s_\A
\non[4pt]
&&{}-\f{\i}{4\sum_\A (J_A^2)}\, \sum_{\A,\Ab,\Ac}
\e^{\A\Ab\Ac}\, J_\Ab J_\Ac \,(w_\Ab^2-w_\Ac^2)\,
\f{w_\A(\l_k\!-\!\l_l)}
{w_\Ab w_\Ac\!\left(\ft12(\l_k\!-\!\l_l)\right)} \;\s_\A \non[6pt]
&&{}+\cO(\l-\l_k) \;.\nn
\eaa
As it turns out, in (\ref{HHH}b) all terms linear in the residues $A_j$
cancel in a way completely analogous to (\ref{HHH}a) shown above. We
hence restrict to the remaining terms. On the l.h.s.~we find making
repeated use of \Ref{div} and \Ref{detSG}:
\baa
\f12\,\res_{\l=\l_k}\; \tr\left[ F^{-1}\,\f{dF}{d\l}\right]^2 &=&
-\f1{8\,\left[\sum_\A (J_A^2)\right]^2}\,\sum_\A
J_A^2\,\f{w_\A(\l_k\!-\!\l_l)}
{w_\A^2\!\left(\ft12(\l_k\!-\!\l_l)\right)} \non
&&{}+\f1{2\,\left[\sum_\A (J_A^2)\right]^2}\,
\sum_{{(\A\Ab\Ac)=(123)}\atop{\rm cyclic}}
J_\Ab^2\,J_\Ac^2\;\f{(w_\Ab^2-w_\Ac^2)^2\,w_\A(\l_k\!-\!\l_l)}
{w_\Ab^2 w_\Ac^2\!\left(\ft12(\l_k\!-\!\l_l)\right)} \non[8pt]
&=& 
-\f1{2\,\left[\sum_\A (J_A^2)\right]^2}\,
\sum_\A J_A^4 \, w_\A(\l_k\!-\!\l_l) \non
&&{}-\f1{2\,
\left[\sum_\A (J_A^2)\right]^2}\,
\sum_{{(\A\Ab\Ac)=(123)}\atop{\rm cyclic}} J_\Ab^2\,J_\Ac^2\;
\f{w_\Ab w_\Ac\!\left(\ft12(\l_k\!-\!\l_l)\right)}
{w_\A\!\left(\ft12(\l_k\!-\!\l_l)\right)}  \non[6pt]
&=&\f12\,\f{\p}{\p\l_k}\,
\ln\left(w_1w_2w_3\!\left(\ft12(\l_k\!-\!\l_l)\right)\right)
\la{Hkr}\\
&&{}-
\f1{\sum_\A (J_A^2)}\,\sum_\A
J_A^2\,\f{\p}{\p\l_k}
\ln\left(w_\A\!\left(\ft12(\l_k\!-\!\l_l)\right)\right) \;.\nn
\eaa
The first term in \Ref{Hkr} obviously cancels against the derivative
of the explicit factor $w_1w_2w_3$ in (\ref{HHH}b). To see the origin of
the second term we note that instead of \Ref{inst} the $\l_k$
derivative of $J_\A$ is given by
\ben
2\sum_\A\f{\p J_\A}{\p\l_k}\, 
w_\A\!\left(\ft12(\l_k\!-\!\l_l)\right) \s_\A ~=~ 
-\f{\p}{\p\l_k} 
\Big\{ G^{\vphantom{-1}}\,\s_3\, G^{-1}\Big\} 
- 2\,\sum_\A J_\A\, 
\f{\p}{\p\l_k} w_\A\!\left(\ft12(\l_k\!-\!\l_l)\right) \;.
\een
The additional term on the r.h.s~which has no linear dependence on the
residues $A_j$ coincides precisely with the second term in
\Ref{Hkr}. Thus, we have shown (\ref{HHH}b).
\bigskip

To finally prove (\ref{HHH}c) we first note that
\baa
\sum_j \res_{\l=\l_j} \,\tr\left[ F^{-1}\,\f{\p F}{\p\mu} 
\;F^{-1}\,\f{dF}{d\l}\right] 
&=& - \f{1}{2\pi\i} \oint_{\p E} \tr\left[ F^{-1}\,\f{\p F}{\p\mu}\; 
F^{-1}\,\f{dF}{d\l}\right]\;d\l \non
&=& -\f{1}{2\pi\i} \oint_a \tr\left[  
F^{-1}\,\f{dF}{d\l}\right]^2\;d\l \;, \la{help}
\eaa
where by $\p E$ we denote a closed path encircling
all the singularities. To show the second equality in \Ref{help}, note
that the integrand is single-valued with respect to shifts along the
$a$-cycle, whereas upon tracing along the $b$-cycle it has the
additive twist 
\ben
\tr\left[ F^{-1}\,\f{\p F}{\p\mu} 
F^{-1}\,\f{dF}{d\l}\right](\l+\mu) ~=~
\tr\left[ F^{-1}\,\f{\p F}{\p\mu} 
F^{-1}\,\f{dF}{d\l}\right](\l) 
-\tr\left[ F^{-1}\,\f{dF}{d\l}\right]^2(\l) \;.
\een
The closed integral along $\p E$ thus reduces to the integral over the
additive twist of the integrand along the $a$-cycle. Equation
\Ref{help} may be used to compute the variation of $H_\mu$ as
\baa
\hH_\mu-H_\mu &=& 
-\f{1}{2\pi\i} \oint_a
\tr\left[ F^{-1}\,\f{dF}{d \l}\, A\;\right]\; d\l-
\f{1}{4\pi\i} \oint_a  
\tr\left[ F^{-1}\,\f{dF}{d\l}\right]^2\;d\l \non[6pt]
&=& 
\sum_j \res_{\l=\l_j} \,
\tr\left[ F^{-1}\,\f{\p F}{\p \mu}\, A\;\right]\; 
+ \f12\,\sum_j \res_{\l=\l_j} \,
\tr\left[ F^{-1}\,\f{\p F}{\p\mu} \;
F^{-1}\,\f{dF}{d\l}\right]^2 \;.
\nn
\eaa
By some further calculations similar to the one given in the proofs of
(\ref{HHH}a) and (\ref{HHH}b), this variation can now be shown to
coincide with the r.h.s.~of (\ref{HHH}c). We leave the details to the
reader. This finishes the proof of Theorem 4.1.  \qed

\section{Open problems}
In this paper we have extended the construction of elementary
Schlesinger transformations for $\sl(2,\C)$-valued meromorphic
connections from the Riemann sphere to the torus. The induced
transformation of the $\tau$-function of the elliptic Schlesinger
system has been explicitly integrated.

There are several ways for a further extension of these results. We
hope, that the formula \Ref{tauhtau} will give rise to new integrable
chains associated to elliptic curves in a way similar to the
construction of integrable chains from isomonodromic deformations on
the sphere \c{JimMiw81b,JimMiw81c}. For a complete generalization of
the program of \cite{JiMiMoSa80}--\cite{JimMiw81c} to the elliptic
case one should extend the results of this work to higher rank
matrices and higher order poles.

An interesting problem would be the generalization of the notion
of Schlesinger transformations for isomonodromic deformations on
higher genus curves, which seems although rather difficult from
the technical point of view, cf.~\c{Ivan96,LevOls97}.
Already on the torus it is a rather nontrivial problem to extend our
construction to isomonodromic deformations with variable twist
\c{KorSam97} as has been discussed in the text.

In the paper \cite{Koro99} it is explicitly solved a certain class of
Riemann-Hilbert problems on the torus which allows to get a class of
solutions of the elliptic Schlesinger system in terms of Prym
theta-functions.

\section{Acknowledgments}
Two of the authors (D.\,K. and H.\,S.) would like to thank University
of Algarve for kind hospitality while the main part of the work was
done. This work was supported by the grant PRAXIS/2/2.1/FIS/286/94
and the EU contract {\em Integrability, non-perturbative effects, and
symmetry in quantum field theory}, ERBFMRX-CT96-0012.

\providecommand{\href}[2]{#2}\begingroup\raggedright\endgroup

\end{document}